\documentclass{article}
\usepackage{graphicx} 
\usepackage[margin=1in]{geometry}
\usepackage{parskip}
\usepackage{amsmath}
\usepackage{amssymb}
\usepackage[backend=biber, style=numeric, citestyle=numeric]{biblatex}
\usepackage{hyperref}

\title{Anthropic Selection for a Low-Entropy Past}
\author{Brendon Matusch \quad \quad \quad \texttt{matusch@stanford.edu}}
\date{}

\begin{document}

\maketitle

\begin{abstract}
The definition of thermodynamic entropy is dependent on one's assignment of physical microstates to observed macrostates. As a result, low entropy in the distant past could be equivalently explained by selection of a particular observer. In this paper, I make the case that because we observe a low-entropy past everywhere even as we look further and further away, anthropic selection over observers does not explain the non-equilibrium state of the observed cosmos. Under a uniform prior over possible world states, the probability of a non-equilibrium past, given our local observations, decreases to zero as the size of the world tends toward infinity. This claim is not dependent on choice of observer, unless the amount of information used to encode the observer's coarse-graining perception function scales linearly with the size of the world. As a result, for anthropic selection to choose a world like the one we live in, the initial state of a universe with size $N$ must be low-information, having Kolmogorov complexity that does not scale with $N$.
\end{abstract}

\section{Introduction}

The universe's increasing entropy is the source of the clear arrow of time we observe, and the resulting distinction between past and future. However, entropy does not exist at the level of fundamental physics. Entropy is an inherently observer-dependent phenomenon originating from incomplete observation of the world around us, opening the possibility that our observed arrow of time could reflect not the world itself but the way we perceive it.

The concept of entropy is defined in many different ways. Shannon entropy is defined as the time-averaged negative log of state probabilities. Thermodynamic entropy is generally defined as the time-averaged log of the number of microstates corresponding to a given macrostate. These two concepts are broadly equivalent \cite{ghosh2011entropy}. A corollary of this equivalence is that thermodynamic entropy requires a choice of ``coarse-graining'', or the bucketing of microstates into macrostates, which corresponds to one's arbitrary choice of $p(x)$ in the definition of Shannon entropy.

In \cite{rovelli2015times}, there is a thought experiment, roughly as follows: we have a box full of gas, and we define a subsystem as all the gas particles in the left half of the box at a specific given time. Even if the box as a whole is in a generic state (already at a maximum-entropy equilibrium), this particular subsystem will exhibit increasing entropy as that particular subset of particles gradually expands to fill the box uniformly. Furthermore, such a subsystem exists not just in this particular scenario, but within almost any sufficiently interconnected system.

This paper is structured as follows. First, I highlight a few important differences between the above thought experiment and thermodynamic properties of the universe we observe. Second, I provide a rough computational model for an arbitrary entropy metric that is consistent with our cosmological observations, and describe the limitations of any such metric when the size of the universe is not fixed or bounded. Finally, I argue along these lines that any meaningful entropy metric \textit{requires} that the initial state of the universe is low-information; in a specific sense that is \textit{not} dependent on our choice of observer or coarse-graining.

\section{General Limitations of Anthropic Subsystem Selection}

The implication I take from the above thought experiment is that anthropic selection over possible subsystems of the universe could explain why the early universe has low entropy; rather than the universe inherently starting in a low-entropy state.

The anthropic principle has found important uses in physics, one of the most famous being Weinberg's estimation of the cosmological constant \cite{weinberg1989cosmological}. One of the challenges with anthropic selection generally is that its results are dependent on our assumed prior distribution of possible worlds. We must assume some prior belief over possible worlds, and then apply Bayes' rule to calculate a posterior distribution over worlds we could live in.

In this paper my objective is to address the scenario in which the universe is in a generic (equilibrium) state, and anthropic selection provides for our observed subsystem. Almost equivalently, we can say that the universe is in a randomly sampled state, since the vast majority of possible states are at or near a maximum-entropy equilibrium. This assumption restricts us to a uniform prior distribution over possible world states, because at thermodynamic equilibrium there can be no ``preference'' for one microstate over any other.

\subsection{Requirement of Minimum Entropy}

In the above scenario, if we project back in time beyond the low-entropy state with our selected particles in the left half of the box, we find in all likelihood that they expand to fill the whole box once again; that is, they were corralled into half of the box only ``by chance''. This is analogous to a Boltzmann brain, where we find a complex and surprising arrangement appears unexplained out of the void, but increases in entropy \textit{in both directions of time}.

This problem can be sidestepped by selecting a subsystem that not only increases in entropy from in its initial state, but takes a minimum at that state, or perhaps is not well-defined before that state. This seems to be the case in our own universe; I am not sure whether it is possible for the case of subset of particles in an isolated box of gas, but it may well be possible given other sets of physical laws.

\subsection{Problems with a Uniform Prior}

The second and more significant problem with anthropic selection is its sensitivity to our assumed prior distribution over subsystems. Suppose instead of having a single box of gas for which we select for a low-entropy subsystem, we have 100 such boxes on different planets, each containing particles indexed from 1 to 10000, and we can choose subsystems by writing down the indices of the particles included.

If we assume the hundred boxes are each initialized independently, and we choose a rule that select for low entropy in the first box, the chance of one of the other boxes being in a similarly low-entropy state is close to zero, and further approaches zero as the number of particles increases. By defining the boxes' initial states independently, we have guaranteed this. If the boxes' initial states have shared information, only then could we plausibly have a correlation between the entropy of corresponding subsystems in different boxes.

Note the correspondence between the choice of a subsystem in this thought experiment, and the choice of an subsystem in our own universe. Sampling the universe's state from a uniform prior distribution, as is required if we assume the universe is in a generic state, and then applying anthropic selection for desired local properties, is analogous to selecting a subset of particle indices in the above.

By ``uniform prior'' I do not mean to suggest that positions and velocities are chosen from uniform distributions, but rather that the world state is selected from among its possible states using a uniform distribution. We can equivalently assume that the universe starts in near-maximum entropy (generic) state.

Of course this is a crude example, but I will argue that it is a useful analogy for our own universe, where we observe many different regions all exhibiting a low-entropy past. Using anthropic selection over a uniform prior, we would expect to find only a \textit{locally} low-entropy past, while in fact we observe it globally. In the rest of this paper I will make that claim more precise.

\section{Subsystem Selection Using a Plausible Entropy Metric} \label{uniform_prior}

Moving on from the thought experiment, I would like to now enumerate the consequences of this type of anthropic selection as an explanation for the observed increasing entropy of our universe. As discussed, the prior distribution over initial states is extremely important.

Suppose we apply anthropic selection to select such an increasing-entropy subsystem out of a maximum-entropy universe; or analogously, we select for a universe containing globally increasing entropy out of a uniform distribution over possible universes. In both cases, we would expect increasing entropy within our own bodies and perhaps biosphere and planet, but an entire universe with a low-entropy past originating from the big bang is unnecessary for our own existence, and therefore either (a) calls out for an explanation more fundamental than mere anthropic selection, or (b) suggests the underlying distribution over world states is not uniform.

We observe similar thermodynamic properties everywhere we look in the universe. If our observed universe is truly an anthropically selected subsystem, the property of such a subsystem being found decreases exponentially with the size of our observed universe. Anthropic selection is not a satisfactory explanation for this because it selects only for our immediate surroundings.

We could define some underlying distribution over possible universe states, such that applying anthropic selection for increasing entropy within our immediate surroundings would also imply increasing entropy for the universe as a whole. I find this a plausible and intuitive explanation; however, it would certainly violate our assumption of a generic maximum-entropy underlying world, in which selection for low entropy of one component does not imply low entropy in another component.

Let us make this problem more concrete by defining some minimal specifications for an entropy metric that increases over time, and examining the consequences of applying anthropic selection to choose an increasing-entropy system.

\subsection{General Properties of an Entropy Metric}

If we have a universe where it is possible to anthropically select a subsystem with increasing entropy, then it must be possible to mathematically write a measure of that subsystem's entropy. Consider a camera attached to a computer which calculates the entropy of the captured image. The camera measures only the entropy of a small and specific subsystem, but it is possible to write a program to compute that entropy from the world state. In essence, the camera is a real-world implementation of that program.

In this conception of the problem, the entropy calculation is modeled as a function of the world state computed from outside the world, rather than the workings of a device or biological brain within the world. However, in this paper we are concerned only with setting asymptotic lower bounds for the complexity of an entropy measure, so this simplification is acceptable, and will include a superset of entropy measures that could actually be physically implemented.

In this section, I will define what we mean when we say that our universe has increasing entropy, and then examine the consequences of using anthropic selection to make this true.

Let us make the following assumptions:

\begin{enumerate}
    \item The world state $X$ at any given time can be written an $N$-bit discrete number, such that there are $2^N$ possible states.
    \item The universe has a deterministic transition function $T(X) = X'$ which takes us from our current state to the next state.
    \item We have conservation of information, meaning our transition function is invertible; so we can also write $T^{-1}(X') = X$.
\end{enumerate}

Of course, these assumptions are not exactly correct (the notion of a slice in time is not well-defined; the world may not be deterministic; we really should be using qubits; etc.) but this is a useful simplified model, and I will argue that the conclusions derived will hold nonetheless.

Given these assumptions, there is no fundamental entropy of the world state, because there is no uncertainty. If we know the entire world state, we can predict it forward and backward perfectly. As in the real world, entropy is instead defined by a coarse-graining function.

\subsection{Specific Requirements}

The entropy we observe in the real world has the important property that when it is not already at equilibrium, it increases over time. This is only a probabilistic statement, but as the size of the world $N$ tends towards infinity, the probability of entropy increasing from one state to the next approaches 1.

Let us write our subsystem entropy metric as $H(X)$, an arbitrary function of our entire world state $X$. Our main restriction on $H(X)$ is that it increases over time from an initial state $X_0$:

\begin{equation*}
H(X_0) < H(T(X_0)) < H(T(T(X_0))) < H(T(T(T(X_0)))) < ... < H(T^M(X_0))
\end{equation*}

To reiterate, this is true only with a probability very close to 1 in the real world; however, it is true with certainty as the world size $N$ tends toward infinity.

We must add one extra restriction to our definition of subsystem entropy: $H(T^{-1}(X_0)) \nleq H(X_0)$. In words, this states that either our entropy takes a minimum at $X_0$, or perhaps it is undefined before $X_0$ (as seems to be the case in our universe).

Our definition of $H(X)$ encompasses any machine that could be built in the world (along with many that could not be built) that measures the entropy of a region in the world. We can alternately interpret $H(X)$ as a timer rather than a measure of entropy; a subsystem with increasing entropy is required to build a system such as a clock that can retain traces of the past.

\section{Asymptotic Properties of Our Entropy Metric}

Let us think of how we would encode $H(X)$. We are assuming we know the entire state of the world, so one approach is simply to take the current state of the world, and step it backward by repeatedly applying $T^{-1}$ until we reach some known initial state $X_0$. This is crudely analogous to how the age of the universe is approximately determined by projecting the movement of galaxies back in time.

In general, we must assume that our definition of $H(X)$ contains knowledge about the transition function $T(X)$. Note that the amount of code in $T(X)$, or its Kolmogorov complexity $K(T)$, is $O(1)$, because our transition function is invariant in the size of the world $N$. Our goal in this section is to determine whether the same can be true of our desired entropy function: whether $K(H) = O(1)$.

By ``code'' and ``encode'' below, I am not referring to any specific programming language, but rather to some general Turing-complete system, in which the Kolmogorov complexity of some definition of the function $H$ is used as the theoretical minimum amount of information required to encode it. We will assume that $N$ approaches infinity, so these complexities are written in big-O notation.

Why is $N$ assumed to approach infinity, when the observable universe has a finite size? Because we are assuming that anthropic selection to choose an increasing-entropy universe, and as argued in Section \ref{uniform_prior}, this selection can be applied only locally, while we observe increasing entropy over the entire universe.

\subsection{High-Information Initial State}

We will consider the trivial transition function $T(X) = X + 1$ as above, because it is easiest to reason about. Since $N$ approaches infinity, for our purposes every transition function $T$ with Kolmogorov complexity $O(1)$ is equivalent to any other. A more complex definition of $T(X)$ may require a more complex definition of $H(X)$ to ``undo'' the computations performed, but would not change the reasoning below.

To trivially satisfy the property of increasing entropy, we can take $H(X) = X$. However, this would not exhibit the property of taking a minimum at $X_0$, since stepping further back in time would reveal that $X_0$ does not have any significance as a minimum value, but is an arbitrarily chosen starting point.

However, given this choice of $T(X)$, we could define a perfectly satisfactory $H(X)$ by taking $H(X) = \mathrm{abs}(X - X_0)$, where $X_0$ is the starting state as above. Note that this requires encoding our universe's initial state in the code for our entropy metric. In principle, there is no way around this; any entropy metric is defined with respect to some special low-entropy state(s). Even the entropy of a deck of cards is defined with respect to a special sorted state.

If we have a truly generic initial world state, where we cannot make any assumptions whatsoever about its structure, such an $H(X)$ requires $O(N)$ bits to encode regardless of our choice of $T(X)$. In fact, we cannot do any better than this, because any parameterized definition of $H(X)$ has to have a single minimum, and any parameterization of $H$ has to contain enough information to encode this minimum ($N$ bits).

Because this requires a definition of $H$ whose size scales linearly with the number of bits in the universe $N$, it is not a tractable definition of entropy. A computer that precisely encodes this entropy metric would take up more space than the entire universe.

Of course, there are specific cases where $X_0$, and thus $H$, can be encoded much more efficiently than that; our own universe is such a case. Below I lay out the conditions under which $H$ can be encoded efficiently.

\subsection{Low-Information Initial State}

Our key assumption above is that we are in a generic (uniformly chosen) initial state $X_0$, and we have shown that if this is true, any computational model for an entropy function satisfying basic properties must take a computer larger than the entire universe to encode. Clearly, this is not the case. Though we are not capable of directly computing the universe's entropy, we are perfectly capable of computing the entropy of a smaller (simulated) system following the same or similar laws; and our programmatic definition of entropy (as a function of world state) does not scale linearly in size with the system whose entropy is computing.

By contradiction, it must be the case that $X_0$ is not generic, and must have Kolmogorov complexity far less than its brute-force encoding of $O(N)$ bits, such that our definition of $H(X)$ does not scale with $O(N)$ either. If we assume that the function for computing entropy is invariant in the size of the universe, then we see that the Kolmogorov complexity of $X_0$ is constant with respect to $N$.

There may seem a contradiction in this reasoning: I began the paper with the argument that the increasing entropy we observe may be an artifact of our perception, and have now argued that this observation requires the universe to have a fundamentally low-entropy initial state. Is the second claim not also observer-dependent?

The difference is that the latter claim is defined asymptotically: the claim is not that the universe's initial state requires little information to encode (which is a meaningless claim, unless you arbitrarily choose a specific encoding system), but that the amount of information required to encode it does not scale with its size. This is key to the concept of Kolmogorov complexity: the choice of encoding has no significance asymptotically.

\subsection{Implications}

I have assumed in this paper that the transition function of the universe is deterministic and invertible, and that we as humans perceive a subset of the information in the universe such that the coarse-graining defined by our perception exhibits an increasing entropy. Under these assumptions, I have argued that the Kolmogorov complexity of the initial state of a universe of size $N$ must be $O(1)$.

If selecting via anthropic bias for an increasing-entropy local environment also happens to select for an increasing-entropy cosmos, we would effectively have an ``entanglement'' (not in the quantum-mechanical sense) between the thermodynamic properties of different regions. That would set an upper limit on how much information can be in the universe's initial state.

In principle, this would suggest that, given enough time, the entire state of a universe like ours with size $N$ could be reproduced by a computer whose size becomes far smaller than the universe itself as $N$ increases. If this claim is false, anthropic selection over subsystems would not result in the observed property of a globally low-entropy past.

\section{Conclusion}

I have shown that, under certain reasonable assumptions, within a universe with its initial state $X_0$ uniformly sampled from possible states, anthropic selection for a locally increasing-entropy subsystem does not yield globally increasing entropy like that which we actually observe. To escape this conclusion, we must assume that the universe's initial state $X_0$ must have Kolmogorov complexity that does not increase with the size of the universe.

My argument is formulated in terms of a discrete-state universe with discrete time, with a deterministic and invertible state transition function. This is an imperfect approximation, and in the future I would like to generalize this work to take quantum uncertainty into account and use measures of quantum information in place of Shannon information.

Even if these assumptions are not fully accurate, the broader point should hold: that anthropic selection for locally increasing entropy within a subsystem does \textit{not} select for increasing entropy over the whole subsystem, if we are truly in a generic universe state.

Finally, observe that most of the logic in this paper is not dependent on the variable of interest being entropy. An analogous case can be made for many attributes which we hypothesize could be due to anthropic bias.

In general, anthropic bias works as a justification for a property like increasing entropy, but only if it is a regional phenomenon (applying to only Earth or the solar system, for instance), or if we have reason to believe it is fixed across the cosmos. If physics permits regional variation in a pattern, and we sample from a uniform distribution over world states, the probability of that pattern holding universally (as appears to be the case for entropy) approaches zero as the size of the universe increases.

To escape this conclusion, which contradicts all of our empirical observations, we must assume that the universe's initial state, and as a direct consequence its \textit{current} state, is not uniformly sampled from all possible states, but could be in principle computed from far less information than it appears to contain.

\printbibliography

@misc{rovelli2015times,
      title={Is Time's Arrow Perspectival?}, 
      author={Carlo Rovelli},
      year={2015},
      eprint={1505.01125},
      archivePrefix={arXiv},
      primaryClass={physics.hist-ph},
}

@article{ghosh2011entropy,
   title={Black Hole Entropy: From Shannon to Bekenstein},
   volume={50},
   ISSN={1572-9575},
   url={http://dx.doi.org/10.1007/s10773-011-0859-y},
   DOI={10.1007/s10773-011-0859-y},
   number={11},
   journal={International Journal of Theoretical Physics},
   publisher={Springer Science and Business Media LLC},
   author={Ghosh, Subir},
   year={2011},
   month=jun, pages={3515–3520} }

@article{weinberg1989cosmological,
  title = {The cosmological constant problem},
  author = {Weinberg, Steven},
  journal = {Rev. Mod. Phys.},
  volume = {61},
  issue = {1},
  pages = {1--23},
  numpages = {0},
  year = {1989},
  month = {Jan},
  publisher = {American Physical Society},
  doi = {10.1103/RevModPhys.61.1},
  url = {https://link.aps.org/doi/10.1103/RevModPhys.61.1}
}

\end{document}